\documentclass[pre,aps,preprint]{revtex4}
\usepackage{epsfig}
\begin{document}
\title{Ludwig Edward Boltzmann}

\author{S.~Rajasekar}

\affiliation{School of Physics, Bharathidasan University,
Tiruchirapalli -- 620 024, Tamilnadu, India}
\email{rajasekar@physics.bdu.ac.in}

\author{N.Athavan}

\affiliation{Department of Physics, H.H.~The Rajah's College, Pudukottai
622 001, Tamilnadu, India}


\begin{abstract}
In this manuscript we present a brief life history of Ludwig Edward Boltzmann and his achivements.  Particularly, we discuss his $H$-theorem, his work on entropy and statistical interpretation of second-law of thermodynamics.  We point out his some other contributions in physics, characteristics of his work, his strong support on atomism, character of his personality and relationship with his students and final part of his life.
\vskip 10pt
\noindent{\emph{Keywords:}} Boltzmann; $H$-theorem; entroy; second-law of thermodynamics
\end{abstract}
 
\maketitle

\section{Introduction}
\vskip 10pt
Boltzmann was born on 20 February 1844 in Vienna, Austria. He was born during the night between Shrove Tuesday and
Ash Wednesday. Boltzmann used to say that this was the reason for the violent swing of his mood from one of great happiness to one of deep depression. His father Ludwig George Boltzmann was a taxation officer. His mother was Katharina Pauernfeind. After his birth his parents moved to Wels and then to Linz. He attended a high school at Linz. When he was fifteen his father died while his mother died in 1885.  He studied physics at the University of Vienna. He was awarded a Ph.D. degree in 1866 for his work on the kinetic theory of gases supervised by Josef Stefan.
\begin{figure}[b]
\begin{center}
\epsfig{figure=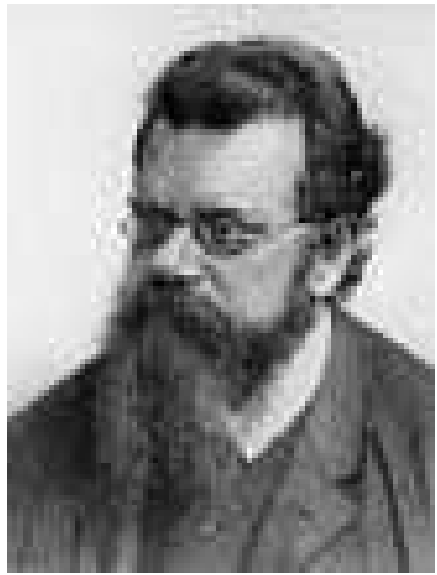, width=0.35\columnwidth}
\end{center}
\caption{Ludwig Edward Boltzmann}
 \label{Fig1}
\end{figure}
\vskip 10 pt
Like molecules, he never stayed very long in any place. He moved from one academic post to another because the polar opposites of his personality gave him no peace. From 1866 to 1869, he worked as a lecturer in Vienna. In 1869 he was appointed as chair of theoretical physics at the  University of Graz. He visited Gustav Kirchhoff and Helmholtz in 1871. He accepted chair of mathematics at Vienna in 1873. Three years after he returned to Graz and became chair of experimental physics. Here he met his wife Henriette von Aigentler. She was the first female student at the University. They had five children, three daughters and two sons. In 1890 he joined as professor at the University of Vienna. In 1893 Stefan died and Boltzmann succeeded him as professor of theoretical physics in his native Vienna. He went to the University of Leipzig in 1900 but returned to Vienna in 1902 because of his dislike of working with Ernst Mach. The Emperor Francis Joseph reappointed him in Leipzig on condition that Boltzmann should not accept a position outside the Empire in future. 
\vskip 10 pt
Influenced by the work of Stefan, Loschmidt, Clausius and Maxwell, Boltzmann began to work on the kinetic theory of gases. He found the energy distribution for gases in 1866. In the year 1871 he formulated the ergodic hypothesis, which formed a basis for the modern version of statistical physics. In the next year, he came up with the famous transport equation. In the same year he derived the $H$-theorem which established a connection between thermodynamics and mechanics. His influence was central to Max Planck's work on black body radiation, Gibbs' 1902 formulation of statistical mechanics and Einstein's work on the light quantum and on Brownian motion. He explained in statistical terms the second-law of thermodynamics. Boltzmann's work received mixed reactions during his lifetime. For years Boltzmann was forced to defend his work against the fairly widespread philosophy of scientific positivism that denied the value of scientific theory and the reality of anything that could be directly observed. Schr\"{o}dinger regarded the Boltzmann line of thought as his first love in science. He said, \emph{No other has ever thus enraptured me or will ever so again}.
\section{$H$-Theorem}
In 1872 he published one of his most important papers containing $H$-theorem and Boltzmann transport equation. This theorem provided the first probabilistic expression for the entropy of an ideal gas.
He introduced probability theory in the context of gas theory. The number of particles in a gas is very enormous and their movements are so swift that one can observe nothing but average values. The calculation of averages is the province of probability calculus. When pointing out what he means by probability he wrote: {\emph{If one wants to buildup an exact theory ... it is before all necessary to determine the probabilities of the various states that one and the same molecule assumes in the course of a very long time and that occur simultaneously for different molecules. That is, one must calculate how the number of those molecules whose states lie between certain limit relates to the total number of molecules}}. 
\vskip 10 pt
He modeled the gas molecules as hard spheres and represented the state of the gas by a time-dependent distribution function of molecular velocities, $f( \overrightarrow{v},t)$. This function represents at each time the relative number of molecules with velocity $ \overrightarrow{v}$. Then with some assumptions finally he arrived at an equation now called \emph{Boltzmann transport equation} which determines the evolution of $f( \overrightarrow{v},t)$:
\begin{equation} \frac {\partial f} {\partial t} + v \frac {\partial f} {\partial x} + F \frac {\partial f} {\partial v} = \frac {\partial f} {\partial t} {\Bigg{\vert}}_{\mathrm{collision}}, 
\end{equation}
where $F$ is a force. The right hand side of the above equation represents the effect of collision.  The above equation completely describes the dynamics of an ensemble of gas particles, given appropriate boundary conditions. He introduced a time-dependent function $H$ as
\begin{equation}
 H =\int f \ln f \ dv, \quad dv = dv_{x}dv_{y}dv_{z}.
\end{equation}
He arrived at the $H$-theorem which states that $H$ can never evolve in an increasing direction. The rate of change of $H$, $dH/dt$, is either negative or zero: $dH/dt$ $\leq$ 0. Negative sign implies that $H$ is decreasing while $dH/dt = 0$ represents equilibrium. That is, $H$ is a negative function always decreasing until the system reaches the equilibrium and $H$ follows the irreversible evolution of a gaseous system. According to him entropy $S$ is proportional to $-H$.
\vskip 10 pt 
Boltzmann's argument appeared to be completely mechanical in nature. Loschmidt argued that the equations of mechanics do not change when $t$ is reversed (replace $t$ by $-t$) implying that physical processes could go backward and forward with equal probability.
\vskip 10 pt
Boltzmann constructed his statistical theory by showing that the probability that a system of molecules with total energy $E$ is proportional to $e^{-hE}$ where $h$ is a parameter function of temperature only. This \emph{Boltzmann factor} evolves in phase space. The probability $dP$ for a system to be in a state represented by points within an element $dw$ is 
\begin{equation}
dP = A e^{-h E} dw, 
\end{equation}
where $A$ is a constant. $ e^{-h E}$ indicates that the probability for occupying a given energy state decreases exponentially with energy. The term $h$ is obtained as $1/kT$. The term $kT$ implies that for a high temperature, it is more probable that a given molecule can be found with energy $E$. Since total probability is one we get 
\begin{equation}
1=\int dP = A \int e^{-h E} dw 
\end{equation}
or 
\begin{equation}
 A = 1 {\Big{/}} \int e^{-hE} dw. 
\end{equation}
Then 
\begin{equation}
 dP =\frac {e^{-h E} dw} { \int {e^{-h E} dw}}. 
\end{equation}
Suppose $S$ is an entropy value and the corresponding probability is $dP$. Then the average entropy $ \overline S $ is 
\begin{equation}
 \overline S = \int S dP = \frac { \int {S e^{-h E} dw}} { \int{e^{-h E}} dw}.
\end{equation}
\section{Entropy and Statistical Interpretation of Second-Law of Thermodynamics}
Boltzmann's most outstanding contribution in physics was the development of classical statistical mechanics to explain how the properties of atoms determine the macroscopic properties and behaviour of the substances of which they were the building blocks. He derived the second-law of thermodynamics from the principles of mechanics.  He argued that the only way to derive thermodynamics from mechanics was to visualize gas as made of atoms. But many great physicists and chemists of his time strongly criticised the concept of atoms. 
\vskip 10 pt
Boltzmann published his statistical interpretation of the second-law of thermodynamics in 1877. Boltzmann wrote this paper in response to objections from Loschmidt who said that the $H$-theorem singled out the direction in time in which $H$-decreases, whereas the underlying mechanics was the same whether time flowed forwards or backwards. In this paper his famous equation  $S=k$ ln $W$, first appeared. Max Planck based the derivation of his black body radiation formula on this equation. This equation connects the entropy $S$ to the logarithm of the number of microstates, $W$, that a given macroscopic state of the system can have, with $k$ now being known as {\emph{Boltzmann's constant}}. According to him entropy must in general increase for an isolated system and showed how such irreversible behaviour could follow from the time-symmetric laws of mechanics.
\vskip 10 pt
Let us now illustrate Boltzmann's notion of entropy. Consider a macroscopic system such as a gas in a box characterized by a set of macroscopic variables $M$. These variables may be volume, temperature and number of particles or molecules. The state of each molecule could be represented by assigning three position components and three momentum components. The state is represented in a six-dimensional phase space. We can alter the positions and momentum of the particles (constituting microscopic states) in several ways without modifying the macrostate $M$. Such alterations of microstates can occur constantly in a gas, that is, without change in the macroscopically measurable properties. Suppose we denote $W$ as  the number of microstates to the macrostate $M$. Boltzmann defined the entropy as $S = k \ln W$, where $k=1.3806505 \times 10^{-23} J K^{-1}$ is the Boltzmann's constant and $W =  N!/ {\prod_{i}} N_{i}!$ where $i$ ranges over all possible molecular conditions and $N$ is the number of identical particles.
\vskip 10 pt
The essential reason for introducing logarithm here is to make the entropy an additive quantity, for independent systems. Thus, for two independent systems, the total entropy of the combined system is the sum of the entropies of each separate system. This is because if the two systems belong to volumes $A$ and $B$, in their respective phase spaces, then the phase space volume for the two together becomes $AB$ since each possibility for one system has to be separately counted with each possibility for the other giving the entropy of the combined system as the sum of the entropies of the individual systems.
\vskip 10 pt
The second-law of thermodynamics states that the entropy of an isolated system not at equilibrium will tend to increase over time approaching a maximum value. The mathematical statement of the second-law is $dS/dt \geq 0$ on average. Essentially, entropy of the universe always increases in any naturally occurring physical process. The greatest achievement of Boltzmann was his explanation for the above. Boltzmann thought that the increase in entropy $S$ of a system is an indication that the particles of the system are essentially moving from a less probable to a more probable arrangement. The state of maximum probability is the equilibrium state and in this state the entropy is a maximum. He assumed that the probability is proportional to number of different complexions, $W$ of the particles, for example, the number of ways in which a total energy $E$ could be divided among $N$ molecules. One can divide the phase space into number of volume elements or cells with all having the same volume. The number of distinct ways of performing an ordered selection of one molecule from $N$ molecules is simply $N$. But the number of distinct ways we can choose two molecules out of $N$ in a specific order is $N(N-1)$. Now selecting $n$ molecules in a specific order is $N!/(N-n)!$. The number of ways of choosing two molecules from $N$ without regard to order is $N(N-1)$ divided by the number of ways two molecules can be ordered, which is 2!. Reasoning along this line the number of ways of choosing $n$ molecules from $N$ molecules without regard to order is $N!/(n!(N-n)!)$. If the cells are numbered as 1,2,\dots$j$, the number of ways of selecting $N_{1}$ molecules from $N$ molecules and placing them in cell $1$, then selecting $N_{2}$ molecules from the remaining $N-N_{1}$ molecules and placing them in cell $2$, \dots and selecting $N_{j}$ molecules from the $N-N_{j}$ molecules and placing them in the cell $j$ is 
\begin{eqnarray}
\omega_{j} & = & \frac {N!} {N!(N-N_{1})!} . \frac {(N-N_{2})!} {N_{2}!(N-N_{1}-N_{2})!}  \dots \frac {N_{j}!} {N_{j}!0!} \nonumber \\
  & = & \frac {N !} {N_{1} ! N_{2} ! \dots N_{j} !}.     
\end{eqnarray}
Then $W$, the total number of distribution is obtained by summing over all the particular distributions $\omega_{j}$. One obtains
\begin{equation}
 S = k \ln \sum_{j}^{} \omega_{j} = k \ln \sum_{j}^{} \frac { N!} {N_{1} ! N_{2} ! \dots N_{j} !}.
\end{equation}
Since the number of molecules in any macroscopic volume of gas is very large we can take the largest term say $w^{*}$, in eq.(2) and neglect the others. Then we get $S = k \ln \omega^{*}$.
\vskip 10 pt
Loschmidt stated that Boltzmann's molecular interpretation of the second-law was in doubt. Boltzmann  replied that his argument was not completely based on mechanics but also on the laws of probability. He proposed that the probability for a certain physical state of a system is proportional to a count of the number of ways the inside of the system can be arranged so that from the outside it looks the same. When the energy quanta is allowed to be infinitesimally small he found $\ln W \propto -H$. Then $S \propto \ln W$.
\vskip 10 pt
Boltzmann's theoretical  argument may appear abstract  and difficult to understand but his conclusion, namely, the entropy-disorder connection is easily to follow. We note that order and disorder and hence entropy are parts of our everyday life. At a temperature molecules in steam are more disorder than those in liquid water. Consequently the entropy of steam is larger than the liquid water. On the other hand, water molecules in the liquid are more disordered than those in ice. Hence, liquid water has more disorder than ice. An ordered pack of cards has low entropy but if they got shuffled the result is the disorder and higher entropy. 
\section{Some Other Contributions in Physics}
In 1879, Josef Stefan experimentally discovered the law which is now known as {\emph{Stefan--Boltzmann law}}. This law states that the total energy radiated per unit surface area of a black body in unit time is directly proportional to the fourth power of its temperature. This law was then theoretically derived by Boltzmann in 1884. Stefan used this law to determine the temperature of the surface of sun. This law can be used to calculate the radius of a star and occurred in the thermodynamics of black holes.
\vskip 10 pt
According to Boltzmann the viscosity of a gas must depend on the mean free path, that is, the average distance a molecule moves between two successive collisions. For hydrogen, its value is about $ 17 \times 10^{-6}$ cm. The frequency of collision is of the order of $10^{9}$ per second. The very high value of the frequency is one reason for the slow diffusion of gases in spite of the great velocity of the molecules.
\vskip 10 pt
He made a significant contribution to the topic of kinetic theory of gases. He formulated the general law for the distribution of energy in a system at a certain temperature. He arrived at the Boltzmann distribution function for the fractional number of particle $N_{i}/N$ occupying a set of states $i$ which each have energy $E_{i}$ is \\
\begin{equation}
 N_{i}/N = \frac {g_{i}} {Z(T)} \exp [-E_{i}/kT],
\end{equation}
where $k$ is the Boltzmann constant, $T$ is temperature, $g_{i}$ is the degeneracy or number of states with energy $E_{i}$, $N$ is $\sum_{i}^{} N_{i}$  and $Z$ is called the {\emph{partition function}} given by $\sum_{j}^{} g_{j} exp [-E_{j}/KT]$. The distribution applies only to particles at a high temperature and low density so that quantum effects can be ignored. In a continuum limit, if there are $g(E) dE$ states with energy $E$ to $E+dE$, the Boltzmann distribution predicts a probability distribution for the energy
\begin{equation}
p(E) dE = \frac {g(E) e^{-E/kT} dE} { \int g(E') e^{-E'/kT} dE'} .
\end{equation}
\vskip 10 pt
The theoretical interpretation of the Dulong Petit rule was given by Boltzmann based on equipartition theorem of classical statistical mechanics. In 1871 Boltzmann showed that the average kinetic energy equals the average potential energy for a system of particles each one of which oscillates under the influence of external harmonic forces. In 1876 he applied these results to a three dimensional lattice. He got the average energy as 3RT= 6 cal/mol. Hence $C_{v}$ the specific heat at constant volume equal 6 cal/mol. deg. His result was in good agreement with experiment for all simple solids except for carbon, boron and silicon.
\section{Characteristics of Work}
A characteristic of his work was that many of his papers were written in great detail. Forbiddingly long with tedious calculations and lacked a clear coherent structure. His 1872 paper on {\emph{$H$-theorem}} and the {\emph{Boltzmann equation}} is about 87 pages long. In 1872 he wrote to his mother that hardly anyone was able to follow him-apart from Helmholtz. In 1873, Maxwell wrote to his colleague Peter Tait: {\emph{By the study of Boltzmann I have been unable to understand him. He could not understand me on account of my shortness and his length was and is an equal stumbling block to me. }}
Once Felix Klein of G\"{o}ttingen asked Boltzmann to contribute an article for the encyclopedia. He hesitated for a long time. Finally Klein wrote to him if he did not do that, he would ask Zermelo to write the article. Zermelo was one among the physicists who opposed the views of Boltzmann. Immediately Boltzmann accepted the assignment.
\vskip 10 pt
He pursued many lines of thought. Jos Uffink wrote: {\emph{He would follow a particular train of thought that seemed promising and fruitful, only to discard it in the next paper for another one, and then pick it up again year later... it makes it hard to pin down Boltzmann on a particular set of rock-bottom assumptions, that would bring his true colours in the modern debate on the foundations of statistical physics.}} 
\section{Strong Supporter of Atomism}
In his period any attempt at all to account macroscopic phenomena in terms of underlying microscopic process was regarded as suspicious. Some physicists and chemists began in search of atomic explanations, while theologically oriented philosophers and positivists dug into preserve their concepts of that atoms were not really {\emph{real}}. Those advocating the concept of atoms were labeled {\emph{materialists}} by classical philosophers. By introducing and obtaining the expression for the Boltzmann factor, Boltzmann became a strong supporter of the reality of atoms. Opposition to his ideas was formidable. Many scientists misunderstood Boltzmann's ideas without grasping the nature of his reasoning. In his lectures on kinetic theory of gases he told his students how much difficulty and opposition he had encountered and how he had been attacked from the philosophical side.
\vskip 10pt
He defended atomic view of matter which was disapproved by Ernst Mach and Ostwald. Jacob Bronowskii wrote: {\emph{Who could think that, only in 1900, peoples were battling, one might say to the death, over the issue whether atoms are real or not. The great philosopher Ernst Mach in Vienna said, `No'. The great chemist Wiehelm Ostwald said, `No'. And yet one man, at the critical turn of the century, stood up for the reality of atoms on fundamental grounds of theory. He was Ludwig Boltzmann ...}}. However, Boltzmann and Ostwald remained close friends and in fact in 1902 Ostwald took steps to appoint Boltzmann in his University in Leipzig. Mach's objections to the atomic theory were of basically a philosophical nature. He considered atoms as something like mathematical functions.
\vskip 10 pt
Ostwald thought that atoms were figments of the mathematics and the energy concept could be useful to formulate grand scheme. George Helm also supported Ostwald's belief. Since no direct evidence for the existence of atoms and molecules is known at that time Ernst Mach opposed Boltzmann's ideas based on atoms. In a meeting in 1890, Ostwald attempted to convince Boltzmann the superiority of his energetics theory over atomism. Boltzmann boldly and suddenly replied that he saw no reason why energy should not be regarded as divided atomicity. In 1895 at a meeting in L\"{u}beck, W.~Ostwald in his presentation of a talk entitled  {\emph{The superseding of scientific materialism}} stated:  {\emph{The actual irreversibility of natural phenomena thus proves the existence of processes that cannot be described by mechanical equations and with this the verdict on scientific materialism is settled}}. Ostwald believed that atoms could not be measured and even if Boltzmann had a mathematical language to juggle pictures of matter, this picture of reality was not to be taken as truth. He further argued that to relate realities, demonstrable and measurable quantities, to each other so that when some are given, the others may be deduced, that is the task of science. Ostwald did not want to accomplish this task by hypostatizing some hypothetical picture, but only by proving relations of mutual interdependence between measurable quantities. Sommerfield who attended the L\"{u}beck meeting said:  {\emph{... The struggle between Boltzmann and Ostwald resembled the battle of the bull against the flexible matador. But this time, the bull won the matador in spite of all his fighting skill. Boltzmann's arguments penetrated. We, all the younger mathematicians, stood at Boltzmann's side}}.
\vskip 10 pt
Loschmidt and Zermelo raised serious criticism on Boltzmann's work. Loschmidt was Boltzmann's former teacher and later colleague at the University of Vienna and a good friend to him. Loschmidt is famous for his estimation of size of atoms. His prime objection was on the prediction of Boltzmann that a gas column in thermal equilibrium in a gravitational field has the same temperature at all heights. 
\vskip 10 pt
In 1904, at the world's fair in St.Louis, Boltzmann told: \emph{No physicist today believes atoms are indivisible}. But the work of Einstein and Perrin brought the end of the war against atoms and molecules. Albert Einstein in 1905 and Jean Perrin in 1908 carried out detailed studies of Brownian motion, the random movement of particles suspended in a liquid, visible under the microscope. They related the movement of suspended particles to the number and energy of the molecules in the liquid which were hitting them. They found a value for the number of molecules in a mole that was consistent with experiments. The direct relationship between the heat energy of atoms and the mechanical energy of Brownian particles provided complete evidence to Boltzmann's interpretation of thermodynamic laws. Einstein stated that molecules of a certain kind could be seen, counted and followed. But Boltzmann was unaware of the theory of Einstein. Finally, Ostwald accepted the concept of atom and in 1909 he wrote in his outlines of general chemistry: {\emph{I am now convinced that we have recently become possessed of experimental evidence of the discrete or grained nature of matter which the atomic hypothesis sought in vain for hundreds and thousands of years}}. 
\section{Character of Personality and Relationship with Students}
He was always very soft hearted but suffered from an alternation of depressed moods with elevated, expansive or irritable moods. His physical appearance being short and stout with curly hair seemed to fit his personality. His fiancee called him her {\emph{sweet fat darling}}. He was a great lover of music and was a talented pianist. He was a great teacher and his lectures were lively, clear and fascinating. He often used stimulating anecdotes. His students were at liberty to ask questions and even to criticize him. He never played up his superiority. He was against dogmatism and made every effort to remove dogmatic views in both scientific and philosophical thoughts. He was open and informal with his students, very sensitive to their needs and aroused admiration and affection on his students. When offering his mechanics course to students in 1902, he said: {\emph{everything (of the course) I have. ... myself, my entire way of this thinking and feeling  ...  strict attention, iron discipline, tireless strength of mind. But forgive me if I ask you that which means most to me: for your trust, your affection, your love in a world, further most you have the power to give, yourself}}.
\vskip 10 pt
According to Lise Meitner, his relationship to students was very personal. He not only saw their knowledge of physics but tried to understand their character. Formalities meant nothing to him, and he had not reservations about expressing his feelings. The new students who took part in the advanced seminar were invited to his house from time to time. There he would play for the students--he was a very good pianist. He shared his personal experiences with his students.
\section{Final Part of His Life}
Towards the end of his life his health deteriorated. He had asthma, headaches, poor eye sight, angina pains. His eyes became so weak that he had difficulty in reading and employed an assistant to read scientific articles for him. His wife wrote his manuscript. During 1906 he announced lectures for the summer semester but canceled them because of his nervous condition. With his wife and daughter he visited a place at the Bay of Duino near Trieste on September 6, 1906. When his wife and daughter were enjoying swimming he hanged himself. He had made an earlier suicide attempt when he was at Leipzig. Physicists everywhere were devastated by the news that Boltzmann, in deep depression, had committed suicide.
\vskip 10 pt
 His act was difficult for many physicists to understand. His suicide seems to have been due to factors in his personal life (depressions and decline of health) rather than to any academic matters. Depressed and in bad health, he committed suicide just before experiment verified his work. In a tribute to Boltzmann, Ostwald described Boltzmann as a victim of the immense sacrifices of health and strength demanded of those who struggle for scientific truth. Some physicists ascribed Boltzmann suicide to {\emph{mental instability}}. According to Chandrasekar, he was greatly depressed by the violent attacks made on his ideas by Ostwald and Mach. This made him to commit suicide.
\vskip 10 pt
Sometimes his suicide is attributed to the unrecognition of his ideas  and works. But in 1888 he was offered a most prestigious position in Berlin but he declined. Universities such as Leipzig, Munich and Vienna shown interest to appoint him with the salaries of several professorships. He was awarded various medals and honorary doctorates. Jos Uffink wrote: {\emph{ ... his death occurred at the dawn of the definitive victory of the atomic view in the works of Einstein, Smoluchowski, Perrin etal. adds a further touch of drama to this picture ... there is no factual evidence for the claim.}} 
\vskip 10 pt
There was no strong evidence that Boltzmann was ignored or suffered due to unrecognition of his work by his contemporaries. On the occasion of his $60^{th}$ birthday, $117$ of the world's most prominent scientists from various countries contributed to the festschrift edited by Stefan Meyer. His suicide seems to have been due to factors in his personal life (depressions and decline of health) rather than to any academic matters. Depressed and in bad health, he committed suicide just before experiment verified his work.  
\vskip 10 pt
His celebrated formula $S= k \ln W$ has been engraved on his tombstone. 
\section{Quotations}
\begin{itemize}
\item
O! immodest mortal! Your destiny is the joy of watching the evershifting battle!
\item
Available energy is the main object at stake in the struggle for existence and the evolution of the world.
\item
Which is more remarkable fact about America:  that millionaires are idealists or idealists become millionaires.
\item
Bring forward what is true \\
Write it so that it is clear \\
Defend it to your last breath!
\item
I am conscious of being only an individual struggling weakly against the stream of time. But it still remains in my power to contribute in such a way that, when the theory of gases is again revived, not too much will have to be rediscovered.
\item
The most ordinary things are to philosophy a source of insoluble puzzles. With infinite ingenuity it constructs a concept of space or time and then finds it absolutely impossible that there be objects in this space or that processes occur during this time.... the source of this kind of logic lies in excessive confidence in the so-called laws of thought.
\end{itemize}
 \vskip 10pt
\noindent{\bf{REFERENCES}}
\renewcommand{\labelenumi}{[\theenumi]}
\renewcommand{\theenumi}{\arabic{enumi}}
\begin{enumerate}
\item
C.~Cercignani, \emph{Ludwig Boltzmann: The Man Who Trusted Atoms} (Oxford University Press, Oxford, 1998)
\label{cerci}
\item
W.H.Cropper, \emph{Great Physicists} (Oxford University Press, Oxford, 2001)
\label{cropper}
\item
R.L.~Sime, \emph{Lise Meitner-A life in Physics} (University of California Press, London, 1996)
\label{sime}
\item
P.~Rife, \emph{Lise Meitner and the Dawn of the Nuclear Age} (Boston: Birkh\"{a}user, 1999)
\label{rife}
\item
JJ O'Connor and E.F.~Robertson, \emph{Ludwig Boltzmann}: http://www-groups.dcs.st-and.ac.uk/~history/Mathematicians/Boltzmann
\label{connor}
\item
S.~Mahanti, Dream 2047 August 2005 
\label{mahanti}
\item
S.~Ramaswamy, Resonance Dec. 2005 176-178 
\label{rama}
\item
Jos Uffink, \emph{Boltzmann's Work in Statistical Physics} in Stanford Encyclopedia of Philosophy  (2004)
\label{uffink}
\item
L.~Boltzmann, Physics Today, January 1992 p.44 (Translated by B.~Schwarzchild)
\label{boltz}
\item
http:$//$ en.wikipedia.org/wiki/Ludwig$_{-}$Boltzmann
\label{wiki}
\item
http:$//$ www.mrs.umn.edu/~sungurea/introdstat/history/w98/Boltzmann
\label{mrs}
\item
http:$//$ www.corrosion-doctors.org/Biographies/BoltzmannBio.htm
 \label{corr}
\end{enumerate}
\end{document}